\documentclass[twocolumn]{aa}
\usepackage{graphicx}
\usepackage{txfonts}

\def\lya{\mbox{Ly$\alpha$}}

\def\lesssim{\mathrel{\hbox{\rlap{\hbox{\lower4pt\hbox{$\sim$}}}\hbox{$<$}}}}
\def\gtrsim{\mathrel{\hbox{\rlap{\hbox{\lower4pt\hbox{$\sim$}}}\hbox{$>$}}}}
\begin{document} 
\newcommand{\magcir}{\ \raise -2.truept\hbox{\rlap{\hbox{$\sim$}}\raise5.truept 
 	\hbox{$>$}\ }}	 
\newcommand{\mincir}{\ \raise -2.truept\hbox{\rlap{\hbox{$\sim$}}\raise5.truept 
 \hbox{$<$}\ }}  

  \title{The Great Observatories Origins Deep Survey} 
  \subtitle{VLT/FORS2 Spectroscopy in the GOODS-South Field: Part III}   
 
     \author{E. Vanzella\inst{1} 
     \and 
      S. Cristiani\inst{1} 
     \and  
      M. Dickinson\inst{2} 
     \and  
      M. Giavalisco\inst{3}   
     \and	
      H. Kuntschner\inst{4}      
     \and  
      J. Haase\inst{4}
     \and
      M. Nonino\inst{1}
     \and
      P. Rosati\inst{5}  
     \\
      C. Cesarsky\inst{5}    
     \and 
      H. C. Ferguson\inst{6} 
     \and
      R.A.E. Fosbury\inst{4} 
     \and
      A. Grazian\inst{7}
     \and
      L. A. Moustakas\inst{8} 
     \and 
      A. Rettura\inst{9}  
     \and
     \\
      P. Popesso\inst{5}
     \and
      A. Renzini\inst{10}  
     \and  
      D. Stern\inst{8} 
     \and
     the GOODS Team 
     } 
 
  \institute{
       INAF - Osservatorio Astronomico di Trieste, Via G.B. Tiepolo 11,
       40131 Trieste, Italy.
     \and
       National Optical Astronomy Obs., P.O. Box 26732, Tucson, AZ 85726.
     \and
     University of Massachusetts, Department of Astronomy, Amherst, MA, 01003
     \and
       ST-ECF, Karl-Schwarzschild Str. 2, 85748 Garching, Germany.
     \and
       European Southern Observatory, Karl-Schwarzschild-Strasse 2,
       Garching, D-85748, Germany.
     \and
       Space Telescope Science Institute, 3700 San Martin Drive,
       Baltimore, MD 21218.
     \and
       INAF - Osservatorio Astronomico di Roma, Via Frascati 33, 
       I-00040 Monteporzio Roma, Italy  
     \and
       Jet Propulsion Laboratory, California Institute of Technology,
       MS 169-506, 4800 Oak Grove Drive, Pasadena, CA 91109
     \and
       Johns Hopkins University, 3400 N. Charles Street, Baltimore, MD, 21218, USA.
     \and
       INAF - Astronomical Observatory of Padova, Vicolo dell'Osservatorio 5, 
       I - 35122 Padova -ITALY 
     \thanks{Based on observations made at the European Southern
Observatory, Paranal, Chile (ESO programme 170.A-0788 {\it The Great
Observatories Origins Deep Survey: ESO Public Observations of the
SIRTF Legacy/HST Treasury/Chandra Deep Field South.}) }
       } 
 
  \offprints{E. Vanzella, \email{vanzella@oats.inaf.it}}

  \date{Received ; accepted }
 
 \abstract
 {}
 {We present the full data set of the spectroscopic campaign of the 
   ESO/GOODS program in the GOODS-South field, obtained with the  
   FORS2  spectrograph at the ESO/VLT.}
 {Objects were selected as candidates for VLT/FORS2 observations 
  primarily based on the expectation that the detection and measurement 
  of their spectral features would benefit from the high throughput and 
  spectral resolution of FORS2. The reliability of the redshift estimates 
  is assessed using the redshift-magnitude and color-redshift diagrams, and
  comparing the results with public data.}
 {Including the third part of the spectroscopic campaign (12 masks) to the previous 
  work (26 masks, Vanzella et al. 2005, 2006), 1715 spectra of 1225 individual
  targets have been analyzed. The actual spectroscopic catalog provides 887
  redshift determinations. The typical redshift uncertainty is estimated to be 
 $\sigma_z \simeq 0.001$.  Galaxies have been 
 selected adopting different color criteria and using photometric redshifts.
 The resulting redshift distribution typically spans two domains:
 from z=0.5 to 2 and z=3.5 to 6.3. 
 The reduced spectra and the derived redshifts are released to the community
 through the ESO web page $\it{http://www.eso.org/science/goods/}$
 \thanks{The catalog (Table~\ref{tab:tblspec}) is available in electronic form
 at the CDS via anonymous ftp to cdsarc.u-strasbg.fr (130.79.128.5)
 or via http://cdsweb.u-strasbg.fr/cgi-bin/qcat?J/A+A/}.
}
 {}
 
 \maketitle
%

\section{Introduction}
The Great Observatories Origins Deep Survey (GOODS) is a public, 
multi-facility project started in the early 2000 designed to gather 
the best and deepest multiwavelength data for studying the formation 
and evolution of galaxies and active galactic nuclei, the distribution 
of dark and luminous matter at high redshift, the cosmological parameters 
from distant supernovae (\cite{riess04}), and the extragalactic background light
(for an overview of GOODS, see \cite{dick01}, \cite{dick03}, \cite{renz03}, 
\cite{giava04a}).

The program targets 
two carefully selected fields, the Hubble Deep Field North (HDF-N) and 
the Chandra Deep Field South (CDF-S), with three NASA Great 
Observatories (HST, Spitzer and Chandra), ESA's XMM-Newton, and a wide 
variety of ground-based facilities. The area common to 
all the observing programs is 320 arcmin$^2$, equally divided between 
the North and South fields. 
In the last five years
the CDF-S has been the target of several spectroscopic campaigns
(\cite{crist00}, \cite{croom01}, \cite{bunk03}, \cite{stan04},
\cite{stro04}, \cite{vanderwel04}, \cite{dick04}, \cite{szo04},
\cite{fevre05}, \cite{mignoli05}, \cite{ravi07}).

This is the third paper in a series presenting the results of the GOODS
spectroscopic program carried out with the VLT/FORS2 spectrograph.
For a full description of its aims we refer to the first
two articles (\cite{vanz05}-RUN1 and \cite{vanz06}-RUN2, V05 and V06, hereafter).
 
Here we recall that the ESO/GOODS spectroscopic program is designed to observe 
all galaxies for which VLT optical spectroscopy is likely to allow the redshift 
determination. 
The program makes full use of the VLT instrument capabilities 
(FORS2 and VIMOS), 
matching targets to instrument and disperser combinations in order to 
maximize the effectiveness of the observations. The magnitude limits 
and selection bandpasses to some extent depend on the instrumental 
setup being used. In the present sample, the mean $z_{850}$ magnitude 
at redshift below 2 is 23.16$\pm$1.20, while for the higher redshift 
part it is 25.34$\pm$0.84. 
 
In the present paper we present the entire FORS2 spectroscopic 
campaign (including new 12 masks). Half of these 12 masks have been 
observed in visitor mode during December 2004 (RUN3, hereafter), and
the other six masks (RUN4, hereafter) in the period 2005 - middle 2006 
in service mode. 
Observations performed in December 2004 were mainly
focused on color-selected Lyman break ``dropout'' targets and the last six masks 
(RUN4) were mainly dedicated to Lyman break ``dropout'' and sources detected 
at 24$\mu$m with the Multiband Imaging Photometer for Spitzer 
(MIPS) instrument aboard the {\it Spitzer Space Telescope}.
180 new redshift measurements out of 345 individual sources 
have been derived from RUN3 and RUN4, several galaxies have been observed two or three times 
(mainly $V_{606}$-dropouts and $i_{775}$-dropouts) in order to gain signal,
in this way 38 confirmation at redshift beyond 5 have been carried out.

The VIMOS spectroscopic survey in the GOODS-S field has completed its run
and hundreds of redshift determinations will be released, mainly in the redshift 
range 0$<$z$\leq$3.5 (Popesso et al. in preparation).

The paper is organized as follows. Sect. 2 describes the target selection,
while Sect. 3 describes the observations and data reductions.
The redshift determination is presented in Sect. 4. In Sect. 5 we discuss
the full data set and in Sect. 6 we discuss two individual sources. 
In Sect. 7 we summarize briefly the whole spectroscopic campaign and give
our conclusions. 

Throughout this paper the magnitudes are given in the AB system 
\footnote{\cite{oke77}, AB~$\equiv 31.4 - 2.5\log\langle f_\nu / \mathrm{nJy} \rangle$}
and the ACS F435W, F606W, F775W, and F850LP filters are designated
hereafter as $B_{435}$, $V_{606}$, $i_{775}$ and $z_{850}$, respectively.
We assume a cosmology with $\Omega_{\rm tot}, \Omega_M, \Omega_\Lambda = 1.0, 0.3, 0.7$
and $H_0 = 70$~km~s$^{-1}$~Mpc$^{-1}$.
 
 
\section{Target Selection}

The selection of galaxies have already been described in V05 and V06. 
Here we recall the adopted criteria:

\begin{enumerate} 
 
\item{Primary catalog:  $(i_{775}-z_{850}) > 0.6$ and $z_{850} < 26$. 
This should ensure redshifts $z \gtrsim 0.7$ for ordinary early-type galaxies
(whose strongest features are expected to be absorption lines),
and higher redshifts for intrinsically bluer galaxies likely to have emission 
lines.}
 
\item{Secondary catalog:  $0.45 < (i_{775}-z_{850}) < 0.6$ and $z_{850} < 26$.}
\item{Photometric-redshift sample:  $1<$ $z_{\rm phot}$ $<2$ and $z_{850} < 26$, 
from \cite{moba04}.} 
 
\item{$B_{435}$, $V_{606}$ and $i_{775}$--dropouts color selected Lyman break 
galaxy candidates (see \cite{giava04b} and \cite{dick04}). The $B_{435}$--dropouts
sources have been selected adopting color criteria slightly different from \cite{giava04b}:
$(B_{435}-V_{606}) > 1.1$ \& $(V_{606}-z_{850}) < 1.6$ \& $(B_{435}-V_{606}) > (V_{606}-z_{850}) + 1.1$ }.
 
\item{A few miscellaneous objects, including host galaxies of
supernovae detected in the GOODS ACS observing campaign.}
 
\end{enumerate} 

In the first RUN (V05), targets were selected from a 
preliminary catalog based on 3-epoch v0.5 GOODS ACS images.

\begin{table*}
\centering \caption{Spectroscopic redshift catalog. $\dag$}
\begin{tabular}{lcccccl}
\hline \hline 
  ID(v1.1)       & $z_{850}$      & $(i_{775}-z_{850})$  & zspec         &class. & Quality & comments \\ 
\hline 
  GDS~J033245.99-275108.3  &     23.48  &0.47   &1.238 &em.    &  B    &[O\,{\sc ii}]3727  \cr 
  GDS~J033246.04-274929.7  &     26.06  &1.77   &5.787 &em.    &  A    &LyA (faint continuum) \cr 
  GDS~J033246.05-275444.8  &     21.49  &0.53   &0.733 &abs.   &  A    &CaH,g-band,H$\beta$,Mg,CaFe \cr 
  GDS~J033246.16-274752.3  &     24.46  &0.43   &1.221 &em.    &  B    &[O\,{\sc ii}]3727  \cr 
\hline   
\multicolumn{6}{l}
{$\dag$ This table is available in its entirety via $\it{http://www.eso.org/science/goods/}$.}\\
\multicolumn{6}{l}
{A portion is shown here for guidance regarding its form and content.}\\
\label{tab:tblspec} 
\end{tabular} 
\end{table*}  
\begin{table}
\centering \caption{Journal of the entire FORS2 observations (RUN1,2,3,4).}
\begin{tabular}{lccc}
\hline \hline
 Mask ID & Date & exp.time (s)\\
\hline
 $RUN1$ (V05) &                   &                \cr
\hline
 990247 &Dec.2002 - Jan. 2003&12$\times$1200\cr 
 984829 &Dec.2002 - Jan. 2003 & 12$\times$1200 \cr 
 985831 &Jan. - Feb. 2003 & 15$\times$1200 + 663 \cr 
 973934 &Jan. 2003&12$\times$1200 \cr 
 952426 &Jan. 2003 & 12$\times$1200  \cr 
 981451 &Jan. - Dec. 2003 & 24$\times$1200 \cr 
 995131 &Oct. 2002 & 8$\times$1800 \cr 
 994852 &Oct. 2002 & 8$\times$1800 \cr 
 990652 &Dec. - Nov. 2002 & 14$\times$1200 + 300 + 900\cr 
\hline
 $RUN2$ (V06)&               &                \cr
\hline
 914250 & Aug. 2003         & 17$\times$1200 \cr
 905513 & Sept. 2003        & 18$\times$1200   \cr
 943018 & Sept. 2003        & 12$\times$1200   \cr
 924345 & Sept. 2003        & 12$\times$1200   \cr
 945143 & Sept. - Oct. 2003 & 12$\times$1200 + 3$\times$1000  \cr
 992438 & Oct. - Dec. 2003  & 12$\times$1200 \cr
 985931 & Nov. 2003         & 12$\times$1200 + 2$\times$120 \cr 
 990204 & Dec. 2003         & 12$\times$1200  \cr 
 904509 & Dec. 2003         & 12$\times$1200  \cr 
 991435 & Dec. 2003         & 12$\times$1200   \cr  
 935030 & Dec. 2003         & 12$\times$1200   \cr 
 951937 & Dec. 2003         & 12$\times$1200 + 1100 + 500   \cr 
 960930 & Dec. 2003         & 12$\times$1200   \cr 
 961839 & Jan. 2004         & 12$\times$1200  \cr 
 932802 & Jan. 2004         & 12$\times$1200   \cr 
 993304 & Jan. 2004         & 12$\times$1200   \cr 
 951526 & Feb. 2004         & 3$\times$1200 \cr 
\hline 
 $RUN3$ &                 &                \cr
\hline
 912940 & Dec. 2004         & 18$\times$1200 \cr
 925109 & Dec. 2004         & 16$\times$1200  \cr
 932249 & Dec. 2004         & 18$\times$1200  \cr
 940129 & Dec. 2004         & 17$\times$1200  \cr
 943544 & Dec. 2004         & 15$\times$1200  \cr
 965910 & Dec. 2004         & 18$\times$1200 \cr
\hline 
 $RUN4$ &                 &                \cr
\hline
 952801 & Nov. - Dec. 2005  & 15$\times$900 + 4$\times$870  \cr 
 952942 & Oct. - Nov.  2005  & 15$\times$900 + 2$\times$870 \cr 
 953015 & Nov. 2005         & 18$\times$900 + 2$\times$870 \cr 
 953048 & Dec. 2005 - Jan. 2006 & 33$\times$900 + 2$\times$870 \cr
 953132 & Feb. - Aug. 2006 & 17$\times$900 + 4$\times$870 + 370 \cr 
 953159 & Jul. - Oct. 2006 & 18$\times$900 + 2$\times$870 + 436  \cr 
\hline
\hline
\label{tab:tblobs} 
\end{tabular} 
\end{table} 

In the following RUNs (2,3 and 4), 
targets were selected from the ACS catalog version r1.1z, based on the 5-epoch v1.0 ACS 
images. The r1.1z catalog is drawn from the r1.0z SExtractor run, 
and merely corrects errors and omissions in the r1.0z catalog files.
The majority of the targets have been selected following the above criteria,
in particular giving priority to LBGs. 
Additionally, in RUN4, sources detected at 24$\mu$m with the 
MIPS were included to fill out the masks.
In addition to requiring 24$\mu$m detection, these sources were selected on their optical
properties to take advantage of the FORS2 red performance, and mainly consisted of 
MIPS-detected extremely red objects (EROs) and distant red galaxies 
(DRGs; see \cite{papo06}) with photometric redshift $z_{phot}< 1.65$, thus putting
them in a range where FORS2 may be capable of measuring redshifts.

For this data release, that includes the entire GOODS/FORS2 campaign 
(RUN1, 2, 3, 4) the objects have been matched to the ACS catalog version 
r1.1z. 

1715 spectra of 1225 individual targets have been extracted
from all RUNs (multiple observations have been carried out,
especially for the high redshift candidates). The FORS2 spectroscopic
catalog lists 887 redshift determinations in the GOODS-S field.

\section{Observations and Data Reduction}

The VLT/FORS2 spectroscopic observations were carried out
in service and visitor (RUN3) mode 
between October of 2002 and October of 2006.  
A summary of the observations is presented in Table~\ref{tab:tblobs}.
Considering the previous works (V05, V06, 26 FORS2 masks) and the
present work (RUN3 and RUN4, 12 masks), a total of 38 FORS2 masks
have been acquired.
In all cases the $300I$ grism was used as dispersing element without  
order-separating filter. 
This grism provides a scale of roughly 3.2\AA~pix$^{-1}$. The nominal 
resolution of the configuration was 
R=$\lambda/\Delta\lambda$=660, which corresponds to
13{\AA} at 8600{\AA}. The spatial scale of FORS2 is $0.126\arcsec$/pixel.
The slit width was always $1\arcsec$. 
In order to effectively improve the sky and fringe 
subtraction and remove CCD blemishes, dithering of the targets 
along the slits was applied, with typically steps 
of 0,$\pm$8 pixels during RUNs 1 and 2, and 0,$\pm$6 pixels in
RUNs 3 and 4. In general the spectral coverage ranges 
from 6000 to 10000\AA.

\subsection{Data Reduction} 

Data have been reduced with a semi-automatic pipeline  
that we have developed on the basis of the MIDAS package (\cite{eso_midas}), 
using commands of the LONG and MOS contexts.
We have used the procedures described in the previous works (V05, V06)
with minor improvements. 

In the case of multiple observations of the same source
in different masks, the one dimensional spectra have been co-added
weighting according to the exposure time, the seeing condition and the
resulting quality of each extraction process (defects present in the CCD, object too
close to the end of the slit, etc.). A visual check of the two
dimensional frames has been performed  and in some cases the two
dimensional spectra have also been co-added, in order to improve and guide the
visual inspection.

We emphasize here that we opted to observe the science targets {\em without} an 
order-sorting filter, implying deleterious effects to the flux calibration.
The second order overlap becomes important at wavelengths above
$\sim$8000\AA~depending on the color of the target.
However, the fluxes derived from spectra are in general
consistent with the ACS photometry (an example for a red and blue source 
is shown in V06, Figure 1).

For the red objects that dominate the FORS2 target selection, we felt that the
improved wavelength coverage more than compensates for the possible
unreliability of the flux calibration. Due to both this second order
light and uncertain slit losses, we caution against using the
calibrated fluxes for scientific purposes.
Fluxes in the released one dimensional spectra are given in units of $10^{-16}$ erg s$^{-1}$
cm$^{-2}$ \AA$^{-1}$.

\section{Redshift Determination}
Spectra of 1225 individual objects have been extracted from all RUNs, 
out of which 887 redshifts have been determined.
In the large majority of the cases the redshift has been calculated through the 
identification of prominent features of galaxy spectra:
depending on the redshift and the nature of the source the 4000\AA\ break, Ca H and K, 
g-band, MgII 2798-2802, AlII 3584, FeII 2344,2383\AA,
Ly$\alpha$, Si\,{\sc ii} 1260.4\AA,  O\,{\sc i} 1302.2\AA, C\,{\sc ii} 1335.1\AA, 
Si\,{\sc iv} 1393.8,1402.8\AA, Si\,{\sc ii} 1526.7\AA, C\,{\sc iv} 1548.2, 1550.8\AA~in 
absorption and Ly$\alpha$, NIV]1485\AA, [O\,{\sc ii}]3727, [O\,{\sc iii}]5007, H$\beta$, H$\alpha$
in emission. 
The redshift estimation has been performed cross-correlating the observed 
spectrum with templates of different spectral types (S0, Sa, Sb, Sc, Elliptical, 
Lyman Break, etc.), using the $rvsao$ package in the IRAF environment.
During the FORS2 campaign we have accumulated spectra of different categories
(LBGs with absorption and emission lines, galaxies in the redshift interval 1.4-2, etc.)
that have been used to build empirical spectral templates to identify similar 
spectral features in the cross-correlating process.

The redshift identifications are summarized in Table~\ref{tab:tblspec} and 
are available at the URL $\it{http://www.eso.org/science/goods/}$. 
 
In Table~\ref{tab:tblspec},  
the column {\em ID} contains the target identifier, that is constructed out of the  
target position (e.g., $GDS~J$033206.44-274728.8) where GDS stands 
for {\bf G}OO{\bf D}S {\bf S}outh.  
The coordinates are based on the GOODS v1.1 astrometry. 
The columns $z_{850}$ and ($i_{775}$-$z_{850}$) list the magnitude 
(SExtractor ``MAG$\_$AUTO'') and the color (SExtractor ``MAG$\_$ISO'') of the sources
derived from the catalog v1.1. The color has been measured through isophotal 
apertures defined in the $z_{850}$ band image (as done in \cite{dick04} and
\cite{giava04b}).

The {\em quality} flag (QF hereafter), indicates the reliability of the
redshift determination. As described in the previous works (V05, V06),
the QF has been divided into three categories: ``A'', ``B'' and ``C''
(secure, probable and tentative redshift determination, respectively).
In general the QF correlates with the goodness of the spectrum. 
In some cases, the presence of defects in the CCD or particularly difficult 
extractions may lead to very noisy one-dimensional spectrum,
however, the presence of evident spectral features in the two-dimensional spectrum
may still give a reliable redshift estimation.
In the FORS2 campaign, 514 sources have been classified with quality
``A'', 226 with quality ``B'', 146 with ``C'', and 349 with ``X'', an inconclusive redshift spectrum.
 
The flag "{\em class}" groups the objects for which emission line(s) (``em.''), 
absorption-line(s) (``abs.'') or both (``comp.'') are detected in the spectrum. 
19 out of 887 sources have been classified as stars.
 
In 28$\%$ of the cases the redshift is based on single emission line, usually 
identified as [O\,{\sc ii}]3727 or Ly$\alpha$.  
In these cases the continuum shape, the presence of breaks, the 
absence of other spectral features in the observed spectral range and the broad band  
photometry are particularly important in the evaluation. The quality for these
sources ranges from ``A'' to ``C'' depending on the additional information 
described above (30$\%$ of the sample with a single emission line have
QF=''A'').
 
The {\em comments} column contains additional information relevant 
to the particular observation. The most common ones summarize  
the identification of the principal lines, the inclination of an emission  
line due to internal kinematics, the weakness of the signal  
(``faint''), the low S/N of the extracted spectrum (``noisy''),  the apparent absence
of spectroscopic lines (``featureless continuum''), etc.
 
In few cases the spectrum extracted is the combination of more than
one source in the slit and where possible the redshifts of the
``components'' have been estimated separately.  In all FORS2 sample, 
11 sources in the GOODS-S field are not present 
in the ACS photometric catalog v1.1.
Six of them have a redshift estimation.  
Three out of six appear to be emission line objects whose continuum is too 
faint and has not been detected in the $z_{850}$ band. The other seven sources 
are outside the GOODS-S area.
\begin{figure} 
 \centering 
 \includegraphics[width=8.7cm,height=9.5cm]{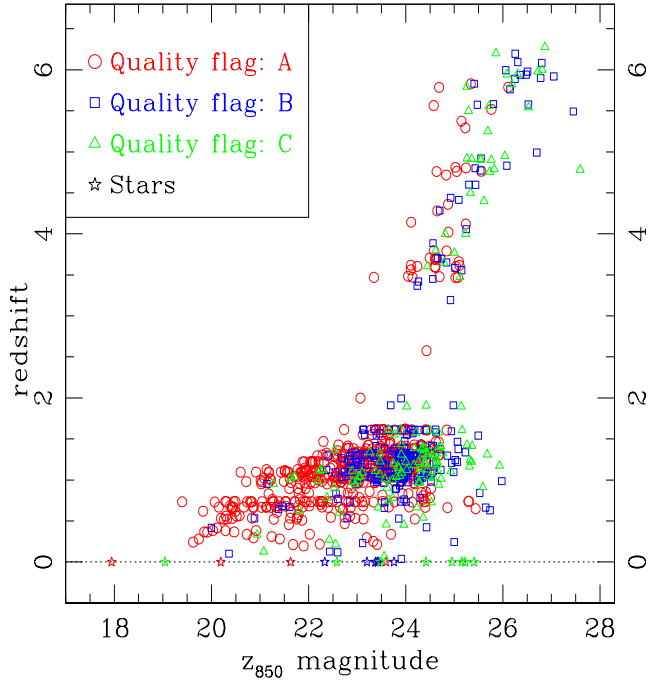}
 \caption{Spectroscopic redshift versus magnitude for the FORS2 catalog
(quality flag ``A'', ``B'', ``C''). Stars are
denoted by star-like symbols at zero redshift.
The gap in the redshift interval 2$<$z$<$3.5 is due
to the spectral coverage adopted ($\sim$ 5800\AA-10000\AA) and will be (partly)
filled with the VIMOS spectroscopic observations.}
\label{fig:z_vs_mag}
\end{figure}
\begin{figure}
 \centering
 \includegraphics[width=8.7cm,height=9.5cm]{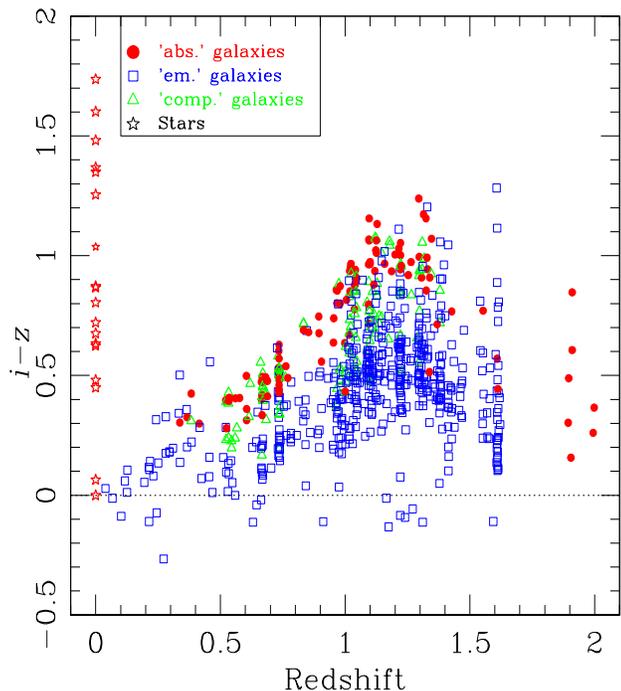}
 \caption{Color-redshift diagram of the spectroscopic sample.
Open red circles are objects identified with absorption
features only (``abs.'' sources),
while open blue squares are objects showing only
emission lines (``em.'' sources).
The intermediate cases are shown by open green triangles (``comp.'' sources).}
\label{fig:i_zVSzspec}
\end{figure}

\subsection{Reliability of the redshifts - diagnostic diagrams}

Figures~\ref{fig:z_vs_mag} and \ref{fig:i_zVSzspec}
show the redshift-magnitude and the color-redshift distributions,
respectively. 

From Figure~\ref{fig:z_vs_mag} it is evident that 
the number of low quality redshift measurements increase 
with increasing magnitude. Also in the case of low redshift galaxies
z$<$0.5 the uncertainty increase due to the adopted spectral range
($\sim$6000 up to $\sim$10000\AA).
In other few cases, serendipitous relatively bright sources enter
in the slit because of the dithering shift, for those sources
the exposure time is reduced and the sky subtraction complicated.
In these cases the quality of the reduced spectra tends to be lower.

Figure~\ref{fig:i_zVSzspec} shows the behavior
of the color-redshift for galaxies at redshift less 
than 2, where the two populations of ``emission-line'' (star-forming)  and ``absorption-line''
(typically elliptical) galaxies are clearly separated.
As expected, the mean color of the absorption-line objects outline the upper 
envelop of the distribution. The intermediate cases ``comp.'' with emission and absorption
features (green triangles) lie in the middle.
The presence of LSS in the field (and discussed below) is evident in the figure
as vertical concentrations of points at redshifts $\sim$ 0.7, 1.1, 1.2, 1.3, 1.6. 

The emission-line objects show in general a bluer $i_{775}-z_{850}$ color and
a broader distribution than the absorption-line sources.
Such bimodality in the color distribution with a larger scatter of  the 
blue component with respect to the red one has been observed in other surveys and can be 
ascribed to the different star formation histories (SFHs) of the two populations. 
The former being the result of a large variety of SFHs that extend to much lower redshift 
than the red ones (resulting in their scattered bluer colors), the latter the result 
of a more similar SFHs in which the bulk of the activity happens at higher redshift 
(\cite{menci05}).

\begin{figure}
 \centering 
 \includegraphics[width=9cm,height=12cm]{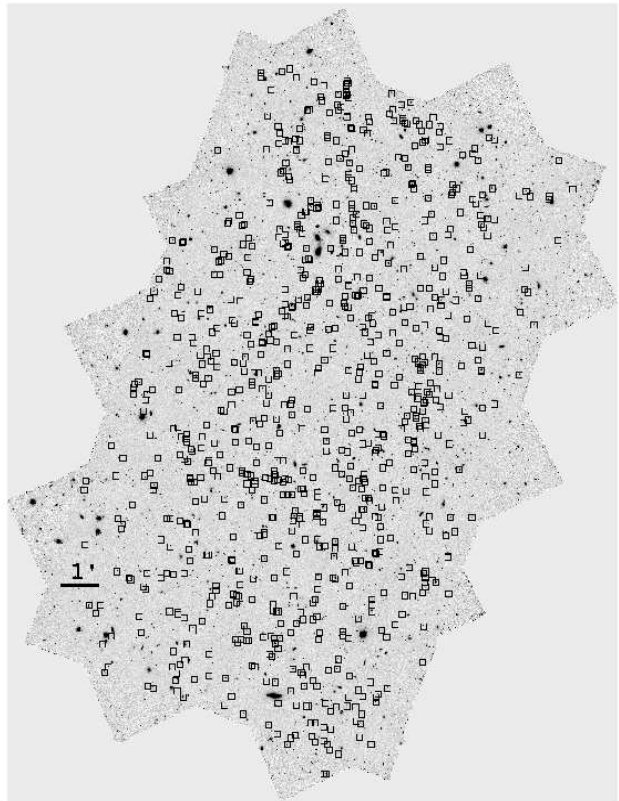}
 \caption{Spatial distribution in the GOODS-S field of the FORS2 spectroscopic catalog 
  (887 redshift determinations).}
\label{fig:spatial}
\end{figure}
\begin{table} 
\centering \caption{Number of sources in the redshift interval 3$<$z$<$6.3 with different spectral 
features.} 
\begin{tabular}{lccc|c}
\hline \hline
redshift range & A & B & C & $N$\\
\hline 
 3 $<$z$\leq$4.5   &  26  & 14 & 11  & 51\cr 
 4.5$<$z$\leq$5.5  &  8   & 9  & 14 & 31\cr 
 5.5$<$z$\leq$6.5  &  5   & 17 & 10 & 32\cr 
\hline
  $N$          &  39  & 40 & 35  & $\bf{114}$ \cr   
\hline 
\label{tab:highz} 
\end{tabular} 
\end{table}

\begin{table} 
\centering \caption{Number of sources in the redshift interval 0$<$z$<$2 with different spectral features 
(without stars).}
\begin{tabular}{lccc|ccc|c}
\hline \hline
 $class$ &$z_{median}$ &$z_{min}$&$z_{max}$& A & B & C & $N$\\ 
\hline 
 emission      & 1.140        & 0.039   & 1.621   & 282  & 152 & 84  & 518\cr 
 absorption    & 1.005        & 0.337   & 1.998   & 86   & 18  & 21  & 125\cr 
 em. \& abs.   & 1.022        & 0.382   & 1.380   & 101  & 9   & 0   & 110\cr 
\hline
  $Total$          & 1.098        & 0.039   & 1.998   & 469  & 179 & 105  & $\bf{753}$ \cr   
\hline 
\label{tab:z_properties} 
\end{tabular} 
\end{table}

\section{Discussion}

\begin{table}
\centering \caption{Completeness of the entire spectroscopic sample describe in the present work
 down to $z_{850}$=26.}
\begin{tabular}{lc|ccc}
\hline \hline
z bin &  zspec & compl.($\%$) & compl.($\%$) & compl.($\%$) \\  
      &        &  $z_{850}<24$&  $z_{850}<25$&  $z_{850}<26$\\  
\hline 
[0.6..0.7[&      46 &11&7&  5\cr  
[0.7..0.8[&      62 &26&18&  13\cr 
[0.8..0.9[&      23 &17&9&  5\cr  
[0.9..1.0[&      54 &22&11&  7\cr 
[1.0..1.1[&      130&48&29&  20\cr
[1.1..1.2[&      85 &77&44&  27\cr
[1.2..1.3[&      107&84&45&  30\cr
[1.3..1.4[&      86 &67&31&  19\cr
[1.4..1.5[&      30 &48&19&  10\cr 
[1.5..1.6[&      15 &37&11&  6 \cr 
[1.6..1.7[&      33 &84&32&  18\cr 
[1.7..1.8[&      0  &0&0& 0  \cr 
[1.8..1.9[&      2  &12&3& 1  \cr 
[1.9..2.0[&      5  &36&8& 3  \cr 
[2.0..2.1[&      0  &0&0& 0  \cr 
\hline 
\label{tab:compl} 
\end{tabular} 
\end{table}

The present release lists 1225 spectra. 887 out of 1225 sources have a redshift 
determination (19 are stars, $z$=0), corresponding to a 
success rate on the redshift determination of about 72$\%$ (assuming that all
redshift measurements are correct).
Sources with inconclusive redshift measurement are in general faint or without
reliable spectral features (indicated in the ``comment'' column 
as $featureless~continuum$ or $faint~continuum$). 
The observations aimed at a uniform sky coverage of the 
GOODS-S field. 
Six FORS2 MXU-masks are necessary to cover the 150 square arcmin of the
GOODS-S field, so with 38 masks (the entire FORS2 campaign) the field has been covered 
roughly $\sim$ 6 times
(the spatial distribution of the 887 sources in the GOODS-S area is shown in Figure~\ref{fig:spatial}).

114 targets out of 887 are confirmed to be LBGs in the redshift range 3$<$z$<$6.3 
(see Table~\ref{tab:highz}), in agreement with the dropout
selection techniques (\cite{giava04b}, \cite{dick03}).

\begin{figure}
 \centering 
 \includegraphics[width=9cm,height=9cm]{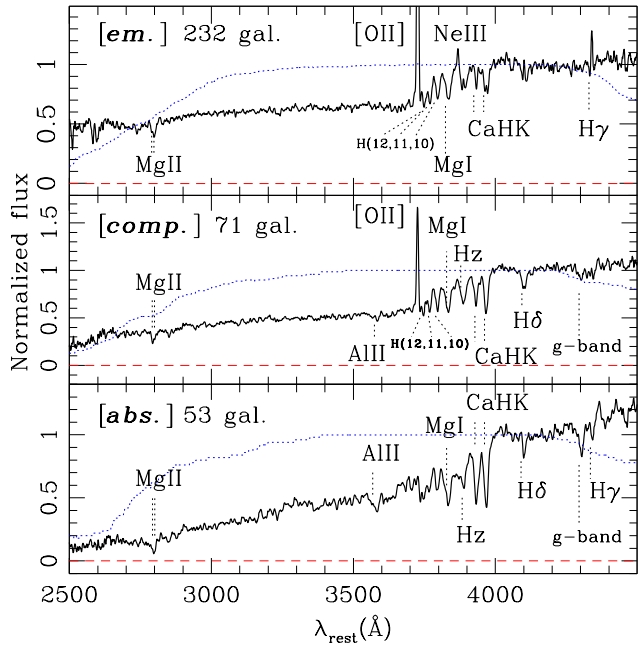}
 \caption{Composite spectra of the three categories ``em.'', ``comp.'' and ``abs.'' from top to
  bottom, respectively (QF=''C'' has not been considered). The median redshift for all is $z\sim$1.
  The co-addition has been weighted and the final spectrum normalized to unity in the wavelength
  range 4000-4100\AA. In the top panel, the [O\,{\sc ii}]3727 line is not fully shown.
  The weight-function is shown as dotted blue line and have also been normalized
  to unity (1 means that all spectra contribute to the sum, 0.5, half of them, etc.).}
\label{fig:stack_lz}
\end{figure}

753 out of 772 sources at redshift less than 2 have been
identified as galaxies (19 are stars, see Table~\ref{tab:z_properties}). 
628 of them (including ``em.'' and ``comp'' classes) show the  [O\,{\sc ii}]3727 
emission up to redshift 1.621, while 125 galaxies identified with absorption lines only
(``abs.'' class, mainly Ca H and K, MgII 2798-2802) span the range of redshift between 0.3-2.0.
The composite spectra of the three categories 
'`em.'' (232 starforming galaxies), ``comp'' (71 intermediate galaxies) and ``abs.'' 
(53 elliptical galaxies) selected in the redshift range 0.8-1.3 
are shown in Figure~\ref{fig:stack_lz}, from top to bottom panels, respectively.

H$\delta$ and g-band are not detected in the ``em.''-like galaxies, but they are well detected in
the other two categories, consistently with the presence of older stellar populations. 
Lines of the Balmer series have been detected down to H12. 
All composite spectra show MgII in absorption.

\begin{figure}
 \centering 
 \includegraphics[width=8.5cm,height=9cm]{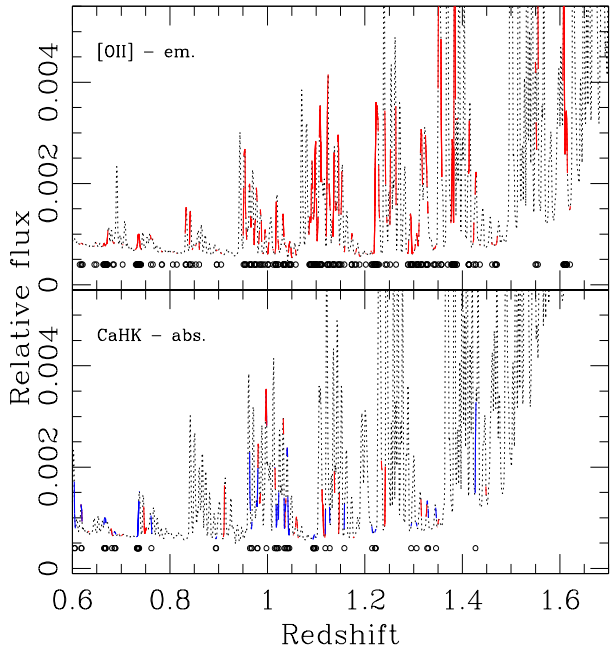}
 \caption{Coverage in the sky emission spectrum by the principal lines used in the
  redshift determination, [O\,{\sc ii}]3727 for emitters (top panel) and Ca H and K 
  for ellipticals (bottom panel, blue K3933.7\AA~and red H3968.5\AA ). Black open circles 
   denote the redshift position in the spectrum.}
\label{fig:sky_lines}
\end{figure}

Obviously, in the presence of emission lines, it is easier to determine a redshift. As an extreme
example, emission lines may be identified without the presence of the continuum in the spectrum
(if the equivalent width is sufficiently large), while absorption lines not.
For this reason the detection of absorption features in relatively faint sources is more uncertain
and more sensitive to the residuals in the sky subtraction, on average they are one magnitude
brighter than the emission lines galaxies (the average $z_{850}$ magnitude
of ``em.''-like galaxies is 23.65 with redshift measurements up to $z_{850}$=26, while
``abs.''-like galaxies have an average of 22.44 and with redshift measurements up to $z_{850}$=25).
A flavour of this effect is shown in Figure~\ref{fig:sky_lines}, where the coverage of the
[O\,{\sc ii}]3727 line (top panel) and Ca H and K absorption doublet (bottom panel) over the
sky emission spectrum is shown as a function of the measured redshift. We note that the
[O\,{\sc ii}]3727 emission has been detected even over the peaks of the strongest sky emission
lines, while the absorption features tend to be identified
in the valleys between the sky lines, in particular beyond redshift 1
(e.g. ellipticals detected at redshift z$\sim$1.3).

This is a likely reason why the majority of galaxies identified in the present
work belong to the ``em.'' class.
Alternatively, or in addition to observational effects, the [O\,{\sc ii}]3727 is a 
classic star forming indicator and the redshift 
interval $1<z<2$ corresponds to the peak of the mean star formation intensity of the universe.
\subsection{Completeness of the sample}

\begin{figure}
 \centering 
 \includegraphics[width=8.5cm,height=9cm]{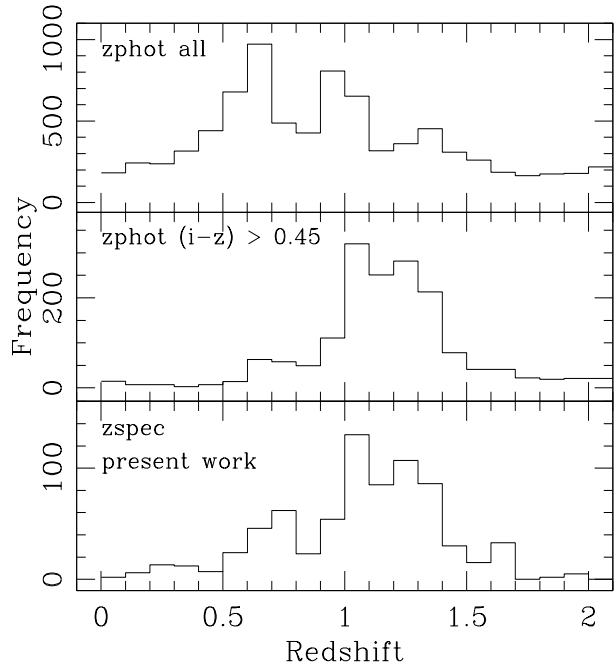}
 \caption{Completeness of the FORS2 spectroscopic sample at redshift less than 2.
  In the upper panel the photometric redshift ($zphot$) 
  distribution down $z_{850}$=26 is shown. In the middle panel, the $zphot$ distribution
  has been plotted adding the main color selection criterium adopted in 
  the FORS2 spectroscopic campaign ($(i-z) > 0.45$, see Sec. 2). In the bottom panel
  the spectroscopic redshift distribution of the present work is shown.
  The values of the completeness are shown in Table~\ref{tab:compl}.
  Photometric redshifts have been drawn from \cite{grazian06}.}
\label{fig:zphot}
\end{figure}

A direct way to derive a rough estimate of the completeness of the present
sample is to compare spectroscopic and photometric redshifts 
down to a common magnitude limit and bins of redshift. 
We have used the recent photometric redshift
measurements in the GOODS-S field, based on the ``PSF-matched'' GOODS-MUSIC 
photometric catalog of \cite{grazian06}. As shown in that work,
photometric redshifts have been estimated with great accuracy,
reaching an average scatter of $\sigma$=0.06.
Table~\ref{tab:compl} summarize the completeness of the 
spectroscopic sample described in the present work.
Column two shows the number of spectroscopic redshifts and columns three, four and five
 show the spectroscopic completeness in each redshift bin (Nzspec/Nzphot) down to
 $z_{850}$=24, 25 and 26, respectively. 
 As targeted, part of the FORS2 campaign has identified galaxies mainly in the 
 redshift range 0.8-2 (see also Figure~\ref{fig:zphot}, middle and bottom panels).
At present, at redshift $\sim$ 1.2 the completeness is $\sim$ 30$\%$
down to $z_{850}$=26.

In the top panel of Figure~\ref{fig:zphot} the photometric redshift distribution
of galaxies in the GOODS-S field with redshift less than 2 and
magnitude $z_{850}$ brighter than 26 is shown. The spectroscopic redshift
distribution is shown in the bottom panel of the same figure, with a 
magnitude limit $z_{850}\sim$26. One of the main target of the FORS2
campaign is to provide spectroscopic identifications in the redshift 
range 0.8-2, according to the color selection criterium $(i-z)>0.45$
(or $(i-z)>0.6$ for the redder ones, see Sect. 2). 
The photometric redshift distribution down to $z_{850}$=26 
of sources with $(i-z)>0.45$ is shown in the middle panel of Figure~\ref{fig:zphot},
in which it is evident that the color cut selects mainly galaxies 
at redshift beyond 1. 
Apart from the normalization, the two distributions (middle and bottom panels) 
show a similar shape. Both the photometric and spectroscopic redshift 
distributions of sources with $(i-z)>0.45$ show approximately 90$\%$ of
galaxies at redshift beyond 0.8. The percentage increases to 96$\%$ if we 
restrict the color cut to $(i-z)>0.6$.
As discussed in V06 and recalled in Sect. 2, galaxies in 
the redshift interval 0.8-2 were selected also on the basis of photometric redshifts 
provided by \cite{moba04}, for this reason the present sample contains confirmed
galaxies at redshift beyond 1 and bluer than $(i-z)=0.45$ 
(see also Figure~\ref{fig:i_zVSzspec}).

\begin{figure} 
 \centering 
 \includegraphics[width=9cm,height=9cm]{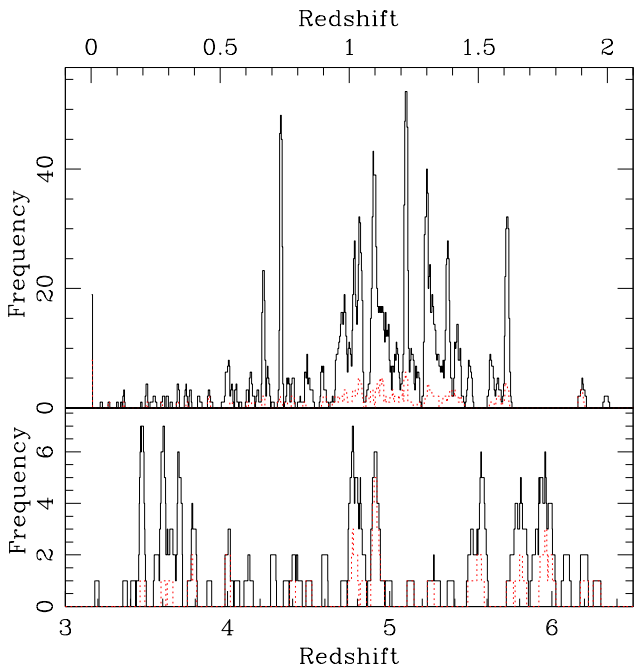} 
 \caption{Redshift distribution according to the selection functions described in Sect. 2
 for the spectroscopic sample with  
 quality A, B (solid line) and C (dotted line). In the top and bottom panels 
 the sources at z$<$2.1 and z$>$3 are shown, respectively. The redshift bin is calculated
 at each redshift and corresponds to a velocity interval $dv$=2000 km/s.}
\label{fig:zdistr} 
\end{figure} 
\subsection{Redshift distribution and Large Scale Structure} 
The top and bottom panels of Figure~\ref{fig:zdistr} show the redshift 
distribution of the galaxies at redshift less than 2 and greater than 3, respectively 
(solid line QF ``A'' and ``B'', dotted line QF ``C''). Galaxies have been counted
in a variable bin $dz$ which has been moved across the redshift interval 
with a step of 0.003 (of the order of 3 times the error of the redshift estimation). 
The bin $dz$ has been calculated at each redshift to be equivalent to a velocity 
$dv$=2000 km/s, so the binning in Figure~\ref{fig:zdistr} is constant in 
the velocity space ($dz$=(1+z)$\times$$dv$/$c$ where $c$ is the speed of light).

In general, the redshift distribution is consistent with the selection criteria adopted,
with the majority of the sources having redshifts in the interval 1$<$z$<$2
and z$>$3 (top and bottom panels of Figure~\ref{fig:zdistr}, respectively).

At redshift less than 2, as discussed in the previous work (V06) and including
the new spectroscopic identifications (RUN3 and 4), large scale structures are confirmed
by the spikes at redshift 0.667 (23 galaxies), 0.735 (49 galaxies), 1.096 (43 galaxies),
1.220 (53 galaxies), 1.301 (40 galaxies), 1.382 (28 galaxies) and 
1.611 (32 galaxies). Some of these peaks are already know in literature
(\cite{cimatti02}, \cite{gilli03}, \cite{fevre04}, \cite{adami05}, \cite{cimatti04}). 
The overdensity of galaxies detected at z$\sim$1.61 is increased by 50$\%$ 
from the previous releases, and counts more than 30 galaxies, all
identified through the [O\,{\sc ii}]3727 emission at $\sim$9730\AA. 

Comparing the roughly expected (photometric)redshift distribution in GOODS-S field 
(middle panel of Figure~\ref{fig:zphot}) and the observed one, 
we note (see Figure~\ref{fig:zdistr}) possible under-densities of galaxies 
in the redshift interval 1-1.5, particularly at 1.06-1.08 and 1.27-1.28. 
The former is close to the sky ``A''-band absorption where the transmission 
is approximately less that 50$\%$ and acts mainly in the [O\,{\sc ii}]3727 
redshift interval 1.03-1.06 (with a less absorbing tail up to 1.074, 
{\it http://tdc-www.harvard.edu/instruments/hectospec/habsky.html}).
This sky absorption band is not introducing an evident bias, 
42 [O\,{\sc ii}]3727 emission line galaxies have been
identified in the more sky absorbed redshift range 1.03-1.06, only 3 galaxies in the redshift interval 
1.06-1.08 (zero sources between 1.06-1.07) and 47 galaxies in the range 
1.08-1.10 where there is the overdensity discussed above.

At redshift beyond 3, galaxies show the expected mean redshift consistently 
with the color selection criteria (\cite{giava04b}, \cite{dick04}). $B_{435}$, $V_{606}$ 
and $i_{775}$-dropouts sources have been confirmed to be at $<z>$ = 3.7, 4.9 
and 5.9, respectively. The success rate of the spectroscopic redshift
measure for these categories is of the order of $70\%$ and the efficiency
of the selection is greater than $80\%$ (Vanzella et al. in preparation).

\section{Individual notes}
\begin{figure}[] \centering
\rotatebox{0}{
  \includegraphics[width=9cm,height=9cm]{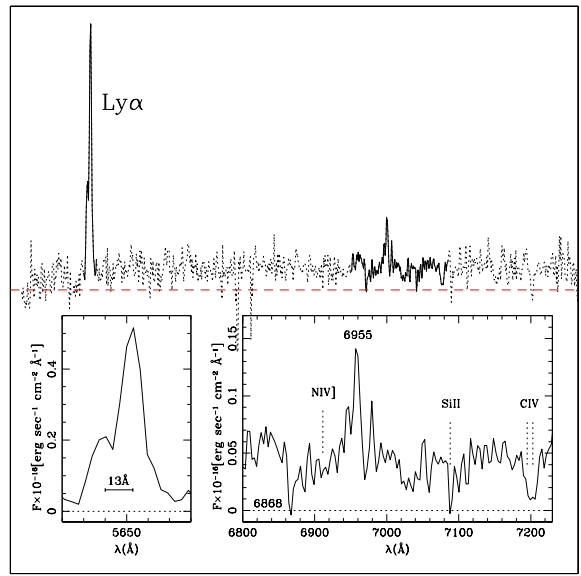}
 }
\caption{One dimensional spectrum of the $B_{435}$-dropout galaxy GDS~J033217.22-274754.4. 
 The bottom panel zoom the solid line regions of the spectrum marked above. 
 The double peaked \lya~line is evident with a separation
 between peaks of $\sim$ 13 \AA. SiII~1526.7 and CIV~1548.2,1550.8 have been detected 
 at redshift consistent with \lya. 
 Other two spectral features have been detected at 6868 and 6955 \AA.}
\label{1dfigure}
\end{figure}
\begin{figure*}[] \centering
  \mbox{
  \includegraphics[width=14cm,height=5cm]{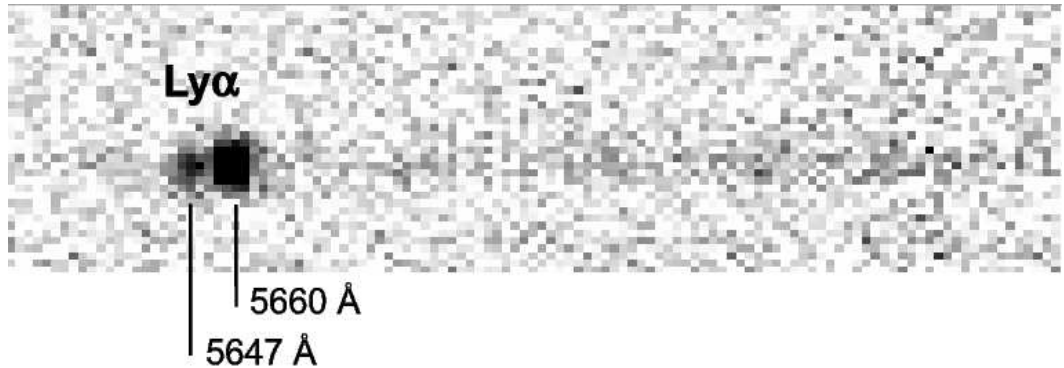}
}
\caption{Two dimensional spectrum of the $B_{435}$-dropout galaxy GDS~J033217.22-274754.4,  
 the wavelength range 5590-5950\AA~is considered. The double peaked line is evident with the 
 continuum redward the structure. The blue spot and the more intense red one
 are separated by 13\AA~and are aligned with the continuum of the target. 
 If one of the two lines is a contamination of the 
 close source GDS~J033217.24-274753.4 at 0.9 arcsec separation from the target, 
 an offset would be expected in the spatial direction. During the observations 
 of this galaxy the average seeing was $\sim$ 0.7 arcsec.}
\label{2dfigure}
\end{figure*}

\subsection{A double peaked Ly$\alpha$ emission line at redshift 3.7}

The source GDS~J033217.22-274754.4 has been 
selected as $B_{435}$-dropout galaxy and has been confirmed to be at redshift
3.649. The main spectral feature is the \lya\ emission line and the
break of the continuum just blueward the line. Two interstellar absorption
lines (SiII~1526.7 and CIV~1548.2,1550.8) have also been detected (see Figure~\ref{1dfigure}).

The emission line at wavelength $\sim$ 5650\AA~shows an evident double peaked structure 
(see Figure~\ref{1dfigure} and Figure~\ref{2dfigure}) with a 
separation of $\sim$ 13\AA. 

Several authors have analyzed the transfer of \lya\ photons in
static and expanding neutral clouds in various physical conditions
(\cite{adams72,urbi81,zm02,ahn03,ver06}). 
If the shell is static, the profile shows two emission peaks with the same
flux, blueshifted and redshifted with respect the systemic redshift.
In the present case the two peaks show different intensities and the
observed separation is much larger than the expected separation produced by
a static shell (0.24\AA~at the redshift of the galaxy), hence, 
the outflow of material seems to be the most plausible reason.

In the outflow picture, the flux of the blue peak is
decreased if the expansion velocity of the shell is increased and
also its position depends to the velocity expansion.
Powered by star formation and supernova explosions,
a wind creates a large shell of swept up material and the observation towards 
such an object would intersect the shell in the front and the 
rear end giving rise to a double peaked emission line profile.
If the expansion velocity is sufficiently high, the blueshifted secondary 
peak disappears and the red wing of the \lya~profile get more flux producing 
the characteristic asymmetric profile observed in high redshift star 
forming galaxies.
\begin{table}
\centering \caption{Lines identified in the spectrum of $B_{435}$-dropouts GDS~J033217.22-274754.4.}
\begin{tabular}{lccc}
\hline \hline
 Line       &  $\lambda$(\AA)   & redshift \cr
\hline  
 & $emission$ & & \cr
\hline              
Ly$\alpha$(1215.8) (blue)& 5647 & 3.645  \cr                      
Ly$\alpha$(1215.8) (red) & 5660 & 3.655  \cr      
??               & 6955 &      \cr
\hline
\hline
 & $absorption$ & \cr
\hline
??           & 6868 &        \cr
SiII(1526.7) & 7089 & 3.643  \cr  
CIV(1548.2-1550.8)  & 7200 & 3.647  \cr
\hline
\hline
\label{tab:lines}
\end{tabular}
\end{table}
Such double peaked profiles of optical
emission lines have been observed for nearby starburst galaxies which exhibit
these large scale outflows typically termed superwinds (\cite{heckman90}).
Additionally, a double peaked \lya\ emission line profile was found from the
starburst galaxy T1214--277 by \cite{mashesse03}, who concluded that
the feature is caused by emission in an outflow.
Double-peaked \lya\ profiles
have been observed also at redshift $z\sim$ 3, in spectra of star forming
galaxies (\cite{fosbury03}, \cite{chri04}, \cite{vene05}, \cite{tap07}).

In the present case, the measured separation between the two peaks
(13\AA) is probably due to an expanding neutral cloud.
Moreover, it is worth to note that the redshift of the blue peak is
consistent with the redshift derived from the SiII~1526.7 and CIV~1548.2,1550.8 lines
(see Table~\ref{tab:lines}). This indicates a plausible connection between
the outflow of the gas deduced from the double peaked \lya\ profile
and that observed from the interstellar lines: 
the blue peak of the \lya\ line and the interstellar absorption 
lines are possibly probing the same behavior of the medium 
(material approaching the observer). 
Interpreting the double-peak as due to an expanding shell, the
velocity difference between the two peaks corresponds to $2~\times~Vexp$ (\cite{ver06}).
The inferred velocity $Vexp$ ($\sim$ 320 km/s) is consistent with the value 
measured by \cite{shapley03} from the velocity difference between stellar and
interstellar lines in a large sample of LBG spectra at redshift $\sim$ 3.

It is worth to note that other two lines have been
detected on the spectrum, one in absorption (6868\AA) and the other 
in emission (6955\AA), that have not been identified.
Detailed spectroscopic observations have to be performed in order to clarify
the nature of such ``additional'' lines.

\subsection{MgII absorbers at z=1.61 in spectra of LBGs at z$>$3 ?}

Two galaxies, GDS~J033226.18-275211.3 and GDS~J033226.76-275225.9 selected as
U and $B_{435}$-dropout, respectively, have been observed 
during the FORS2 and VIMOS campaign (Popesso et al. in preparation).

\begin{figure}[] \centering
\rotatebox{0}{
  \includegraphics[width=9cm,height=9cm]{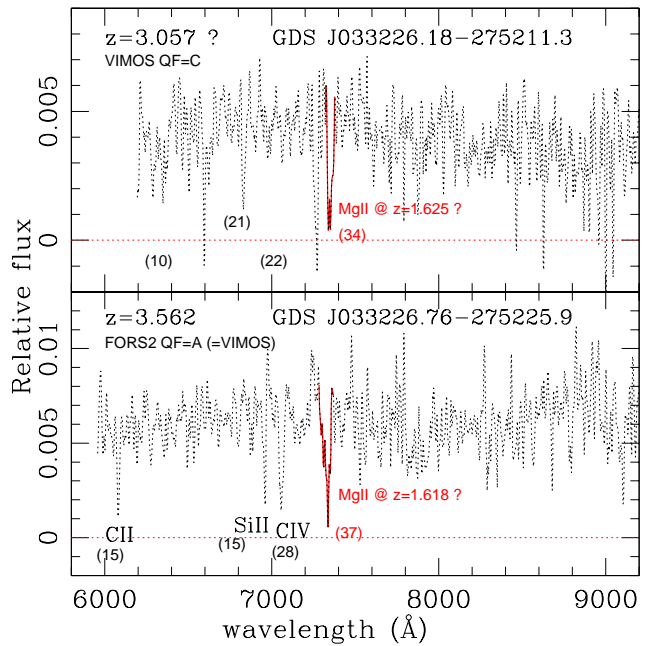}
  }
\caption{One dimensional spectra of two galaxies at redshift 3.057 (top panel)
 and 3.562 (bottom panel). In the bottom case the redshift estimation is
 reliable (FORS2 agree with the VIMOS one). The top source has a VIMOS redshift 
 3.057 with quality ``C'', but no evident correspondence has been found with the 
 absorption feature detected in the FORS2 spectrum. 
 Both spectra show an absorption line at 7349.6\AA~(top) 7331.9 \AA~(bottom), apparently
 not consistent with the redshift of the host galaxy. The FWHM of the lines is shown
 within parenthesis. If interpreted as MgII foreground absorbers (the FWHM of the lines 
 suggest a doublet), they would be at redshift 1.61, at the same redshift of the 
 overdensity discussed above.
 The physical separation at $z\sim$1.61 between them is 140 kpcs.}
\label{IGM}
\end{figure}

In the case of GDS~J033226.76-275225.9 (bottom panel of Figure~\ref{IGM})
the measured redshifts from FORS2 (3.562, QF=''A'') and VIMOS (3.554, QF=''C'') 
agree within the errors. In the FORS2 spectrum, the CII~1335.1, 
SiII~1526.7 and CIV~1548.2,1550.8 have been 
detected. The latter doublet shows a FWHM larger than other single lines
(see Figure~\ref{IGM} where the numbers in the parenthesis indicate the FWHM of the
lines). The absorption line at 7331.1\AA~ may be associated
to FeII~1608.5 in absorption at the redshift of the galaxy. 
However such a feature is unusual if compared with other LBGs, moreover
the FWHM of the line is evidently large (37\AA) with respect to the other lines.

Similarly, in the spectrum of GDS~J033226.18-275211.3 there is an absorption
line with a FWHM that apparently is not consistent with the redshift 
of the source (z=3.057 from the VIMOS spectroscopic survey). The redshift is 
not reliable (QF=''C'') and the FORS2 absorption features (marked with their
FWHM in the top panel of Figure~\ref{IGM}) are not consistent with the VIMOS
redshift. However, the two relatively strong absorption lines around 
7300\AA~seem to have no easy solution if associated to the galaxy. 
We note that also in this case the line at 7349.0\AA~ shows a FWHM of 34.1\AA, 
similarly to the line discussed above.

The fact that the two galaxies are relatively close in the sky (16.5 arcsec)
and show two features at similar wavelength, apparently 
not consistent with the galaxy redshift, support the interpretation 
in terms of intervening absoprtion.
Moreover, in both cases, the FWHM of the lines is large and suggest the presence of a 
doublet (as we see for the CIV~1548.2,1550.8 doublet, bottom panel of Figure~\ref{IGM}).
Interestingly, if the two spectral features are MgII~2798,2802 absorbers, their 
redshifts turns out to be 1.618 and 1.624, very close to the overdensity at redshift 
1.611 detected in the field (see Figure~\ref{fig:zdistr} and Sect. 5.2).

\begin{figure}[] \centering
\rotatebox{0}{
  \includegraphics[width=9cm,height=9cm]{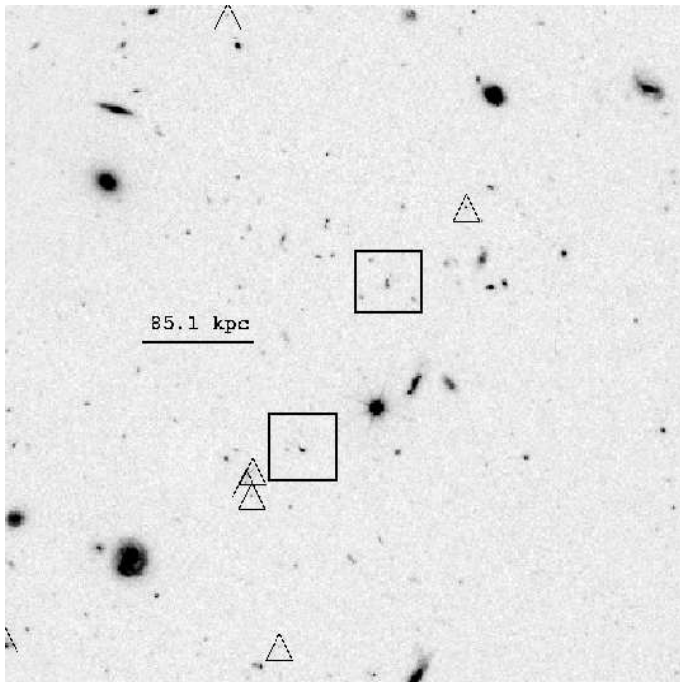}
  }
\caption{z850 ACS image of a portion of the GOODS-S field. The two galaxies 
 at redshift beyond 3 which display possible MgII in absorption at $z\sim$1.61 are 
 marked with squares. The physical separation between them at z=1.61 
 is $\sim$140 kpc (10 arcsec correspond to 85.1 kpc, segment in the figure). 
 Sources with photometric redshift in the interval 1.4-1.8 are
 marked with triangles. The image is one square arcminute size.}
\label{IGM_ACS}
\end{figure}

Unfortunately no confirmed galaxy at redshift 1.61 has been currently 
identified in the neighborhood of the high-z ones (the closest one is at
a physical separation of $\sim$ 900 kpcs from them), further spectroscopic 
investigations would be necessary to shed light on the nature of the absorbers.
The physical separation of the absorbers (140 kpcs) and possible candidate 
members of the overdensity at $z$=1.61 associated with the present 
MgII features are shown in Figure~\ref{IGM_ACS} (triangles indicate 
sources with photometric redshift in the interval 1.4-1.8).
 
\section{Conclusions} 
As a part of the Great Observatories Origins Deep Survey, a 
large sample of galaxies in the Chandra Deep Field South has been 
spectroscopically targeted. 
After the RUN1 (V05), RUN2 (V06) and the present work
(RUN1,2,3,4) a total of 1225 objects with $z_{850} \mincir 27$ have been 
observed with the FORS2 spectrograph at the ESO VLT providing 887 
redshift determinations with a typical $\sigma_z \simeq 0.001$
(16 $\%$ of the sample has a tentative redshift measurement, QF=''C'').
The reduced spectra and the derived redshifts are released to the community  
($\it{http://www.eso.org/science/goods/}$). 
They constitute an essential contribution to reach the scientific goals 
of GOODS, providing the time coordinate needed  
to delineate the evolution of galaxy masses, morphologies, and star 
formation, calibrating the photometric redshifts that can be derived from the 
imaging data at 0.36-8$\mu$m and enabling detailed studies  
of the physical diagnostics for galaxies in the GOODS field. 

Two individual sources have been discussed: 1) a double peaked
\lya\ profile at redshift $\sim$ 3.7 due to outflow of material powered
by star formation and supernova explosions and 2) the possible detection
of two (relatively close, 16.5 arcsec) MgII absorbers at 
redshift 1.61 (the same of the overdensity already
confirmed in the field) observed through spectra of LBGs.

\begin{acknowledgements} 
 We are grateful to the ESO staff in Paranal and Garching who greatly helped 
 in the development of this programme.
 We wish to thank the referee for comments which improved the paper.
\end{acknowledgements}

\end{document}